\documentstyle[12pt]{article}
\begin{document}
\def\be{\begin{equation}}
\def\ee{\end{equation}}
\def\ba{\begin{array}}
\def\ea{\end{array}}
\def\bea{\begin{eqnarray}}
\def\eea{\end{eqnarray}}
\parskip=6pt
\baselineskip=22pt
{\raggedleft{{ ASITP}-94-61\\}}
{\raggedleft{November,1994.\\}}
\bigskip
\bigskip
\bigskip
\medskip
\centerline{\Large \bf $N=2$ and $N=4$ SUSY Yang-Mills action }
\centerline{\Large \bf on $M^4\times (Z_2\oplus Z_2)$ non-commutative geometry
}
\vspace{10ex}
\vspace{10ex}
\centerline{\large  Bin Chen\footnote{e-mail address : cb@itp.ac.cn}\ \ \ \
Hong-Bo Teng\footnote{e-mail address : tenghb@itp.ac.cn} \ \ \ \ Ke
Wu\footnote{e-mail address : wuke@itp.ac.cn}}
\vspace{1.5ex}
\centerline{ Institute of Theoretical Physics, Academia Sinica,
P. O. Box 2735, Beijing 100080, China.}
\vspace{8ex}
\vspace{8ex}

 \centerline{Abstract}
{\it We show that the $N=2$ and $N=4$ SUSY Yang-Mills action can be
reformulated in the sense of non-commutative geometry on $M^4\times (Z_2\oplus
Z_2)$ in a rather simple way. In this way the scalars or pseudoscalars are
viewed as gauge fields along two directions in the space of one-forms on
$Z_2\oplus Z_2$.
}

\newpage

\section{Introduction}

Since A. Connes introduced his non-commutative geometry into particle
physics\cite{connes1,conneslott1}, a successful geometrical interpretation of
the Higgs mechanism and Yukawa coupling has been acquired. In this
interpretation, the Higgs fields are regarded as gauge fields on the discrete
gauge group. The bosonic part of the action is just the pure YM action
containing the gauge fields on both continuous and discrete gauge group, and
the Yukawa coupling is viewed as a kind of gauge interactions of fermions.

At the same time, applying non-commutative geometry to SUSY theories has
encountered many difficulties. A natural way is to introduce a non-commutative
space which is the product of the superspace and a set of discrete points,
similar to those which have been done in non-SUSY theories. However, it proved
that such an extension of superspace is rather difficult to accomplish.
In\cite{alisusy}, A.H. Chamseddine gave an alternative way in which the
supersymmetry theories in their component forms were considered. He showed that
$N=2$ and $N=4$ SUSY Yang-Mills actions could be derived as action functionals
for non-commutative spaces, and in some cases the coupling of $N=1$ and $N=2$
super Yang-Mills to $N=1$ and $N=2$ matter could also be reformulated. This
paper is the first one in which the non-commutative geometry is successfully
applied to SUSY theories.

Intrigued by A.H. Chamseddine's work, we show the $N=2$ and $N=4$ SUSY
Yang-Mills actions can be derived with non-commutative geometry on $M^4\times
(Z_2\oplus Z_2)$ in a rather simple way. We use the differential calculus on
discrete group developed by A. Sitarz\cite{sitarz}. With a specific case of
differential algebra on $Z_2\oplus Z_2$, we take the scalar components of the
super Yang-Mills field as gauge fields on discrete symmetry group. In $N=2$
case, the scalar and pseudoscalar fields are set to two directions of the space
of one-forms. In case of $N=4$, the complex scalar fields which belong to $6$
of $SU(4)$ are regarded as components of the connection one-form.

First in section 2 we show the specific case of the differential algebra on
$Z_2\oplus Z_2$. Then in section 3 we derive the $N=2$ and $N=4$ super
Yang-Mills Lagrangian with this algebra. Finally we end with conclusions.

\section{Differential Calculus on $Z_2\oplus Z_2$}

In this section, we briefly describe the differential calculus on $Z_2\oplus
Z_2$ and introduce a specific case of this algebra. For a detailed description
of differential calculus on discrete group, we refer to \cite{sitarz}. Let's
write the
four elements of $Z_2\oplus Z_2$ as \[ (e_1, e_2),\ \ (r_1, e_2),\ \ (e_1,
r_2),
\ \ (r_1, r_2).\] And the group multiplication is
\be
(g_1, g_2)(h_1, h_2)=(g_1h_1, g_2h_2).
\ee
Let $\cal A$ be the algebra of complex valued functions on $Z_2\oplus Z_2$.
 The derivative on $\cal A$ is defined as
\be
\partial_gf=f-R_{g}f   \hspace{4ex} g\in Z_2\oplus Z_2 ,\hspace{1ex} f\in\cal A
\ee
with  $R_{g}f(h)=f(hg)$. We will write $\partial_i$ and $R_i$ for convenience
where $i=1, 2, 3$ refers to $(r_1, e_2), (e_1, r_2), (r_1, r_2)$ respectively.

$\{\partial_i,\ \ i=1,2,3\}$ forms the basis of $\cal F$, the space of left
invariant vector fields over $\cal A$. One can find that the following
relations hold
\be\ba{lll} \label{alrelation}
\partial_1\partial_2&=&\partial_1+\partial_2-\partial_3 \\
\partial_2\partial_3&=&\partial_2+\partial_3-\partial_1 \\
\partial_3\partial_1&=&\partial_3+\partial_1-\partial_2
\ea\ee
\be\ba{lll}
\partial_1\partial_1&=&2\partial_1 \\
\partial_2\partial_2&=&2\partial_2 \\
\partial_3\partial_3&=&2\partial_3
\ea\ee
\be\ba{lll}
\partial_1\partial_2&=&\partial_2\partial_1 \\
\partial_1\partial_3&=&\partial_3\partial_1 \\
\partial_2\partial_3&=&\partial_3\partial_2 .
\ea\ee

The Haar integral on $Z_2\oplus Z_2$ is defined as
\be
\int_{Z_2\oplus Z_2} f=\frac{1}{N_{Z_2\oplus Z_2}}
\sum_{g\in Z_2\oplus Z_2}f(g)\ \ \ \ f\in {\cal A}
\ee
here $N_{Z_2\oplus Z_2}=4$ is the size of $Z_2\oplus Z_2$.

Next we introduce the space of one-forms, $\Omega^1$, which is the dual space
of $\cal F$. Haven chosen the basis of $\cal F$, we automatically have the dual
basis of $\Omega^1$, $\{\chi^i,\ \ i=1,2,3\}$. It is defined with
\be
\chi^i(\partial_j)=\delta^i_j   \hspace{4ex} i, j=1,2,3.
\ee

The definition of higher forms is natural, $\Omega^n$ is taken to be tensor
product of $n$ copies of $\Omega^1$,
\[\Omega^n=\underbrace{\Omega^1\otimes\cdots\otimes\Omega^1}_{n\ \ copies},\]
and $\Omega^0={\cal A}$. And $\Omega=\sum_\oplus\Omega^n$ is the tensor algebra
on $Z_2\oplus Z_2$.

As usual, the external derivative satisfies the graded Leibniz rule and is
nilpotent
\be\ba{l}\label{a}
d(ab)=d(a)b+(-1)^{deg a}a(db)\ \ \ \ a,b\in\Omega\\[3mm]
d^2=0
\ea\ee
and for $f\in \Omega^0={\cal A}$
\be\label{b}
df=\partial_1f\chi^1+\partial_2f\chi^2+\partial_3f\chi^3.
\ee

{}From (\ref{a}),(\ref{b}), one finds
\be
\chi^if=(R_if)\chi^i \ \ \ i=1,2,3\ \ \ f \in {\cal A}
\ee
and
\be\ba{lll}
d\chi^1&=&-\chi^1\otimes\chi^2-\chi^1\otimes\chi^3+\chi^2
\otimes\chi^3-2\chi^1\otimes\chi^1-\chi^2\otimes\chi^1-\chi^3\otimes
\chi^1+\chi^3\otimes\chi^2  \\
d\chi^2&=&-\chi^2\otimes\chi^3-\chi^2\otimes\chi^1+\chi^3
\otimes\chi^1-2\chi^2\otimes\chi^2-\chi^3\otimes\chi^2-\chi^1\otimes
\chi^2+\chi^1\otimes\chi^3  \\
d\chi^3&=&-\chi^3\otimes\chi^1-\chi^3\otimes\chi^2+\chi^1
\otimes\chi^2-2\chi^3\otimes\chi^3-\chi^1\otimes\chi^3-\chi^2\otimes
\chi^3+\chi^2\otimes\chi^1
\ea\ee

We also need the involution $\bar{\ }$ on our differential algebra, which
agrees with the complex conjugation on $\cal A$ and (graded) commutes with $d$,
i.e. $d(\overline{a})=(-1)^{deg a}\overline{da}$. For one-forms,
\be
\overline{\chi^i}=-\chi^i \ \ \ i=1,2,3.
\ee

If we postulate that the one-forms to be anticommutative, i.e.,
\be\ba{l} \label{pos}
\chi^i\otimes \chi^i=0\\[3mm]
\chi^i\otimes \chi^j=-\chi^j\otimes \chi^i   \hspace{4ex} i,j=1,2,3
\ea\ee
then we get
\be \label{posre}
d\chi^i=0   \hspace{4ex} i=1,2,3.
\ee
It is a specific case of the above algebra. Obviously the postulation induces
an antisymmetric tensor algebra $\Lambda$ which is a subalgebra of $\Omega$. In
this subalgebra, we denote the product of forms with $\wedge$. And we will use
this subalgebra to derive the $N=2$ and $N=4$ SUSY Yang-Mills action in next
section.

\section{$N=2$ and $N=4$ super Yang-Mills action}

Let's first consider the $N=2$ case. The $N=2$ super Yang-Mills Lagrangian is
given by
\be
{\cal L}=\ba[t]{l} Tr(-\frac{1}{4}F_{\mu\nu}F^{\mu\nu}
+\frac{1}{2}D_\mu SD^\mu S
+\frac{1}{2}D_\mu PD^\mu P
+i\bar{\psi}\gamma^\mu D_\mu\psi
+\bar{\psi}[S+\gamma_5P,\psi]\\[3mm]
+\frac{1}{2}[S,P]^2)
\ea
\ee
where $\psi$ is the Dirac $4-$spinor which is the combination of two
$2-$spinors, $S$ is the scalar and $P$ is the pseudoscalar. All fields are
valued in the adjoint representation of the gauge group, i.e.,
\be
f=f^a T^a \ \ \ \ \ \ f=A_\mu, S, P, \psi
\ee
where $T^a$ is the generator of gauge group. And
\be\ba{l}
F_{\mu\nu}=\partial_\mu A_\nu-\partial_\nu A_\mu+i[A_\mu, A_\nu]\\[3mm]
D_\mu S=\partial_\mu S+i[A_\mu, S]\\[3mm]
D_\mu P=\partial_\mu P+i[A_\mu, P]\\[3mm]
D_\mu \psi=\partial_\mu \psi+i[A_\mu, \psi].
\ea\ee

{}From a simple observation, we see that it is possible to choose the
generalized connection one-form on $M^4\times (Z_2\oplus Z_2)$
\be\label{connection}
A(x,g)=iA_\mu(x,g) dx^\mu-iS(x,g)\chi^1-i\gamma_5P(x,g)\chi^2 \ \ \ g\in
Z_2\oplus Z_2.
\ee
And we set
\be\label{assign1}
A_\mu(x,g)=A_\mu(x);\ \ \ S(x,g)=S(x);\ \ \ P(x,g)=P(x);\ \ \ g\in Z_2\oplus
Z_2.
\ee
And later in fermionic sector we will also let $\psi(x,g)=\psi(x)$. Since all
quantities take the same values on different points of $Z_2\oplus Z_2$, we will
ignore the variable $g$. Please note in this assignment the scalar and
pseudoscalar are assigned to two directions of the connection one-form, and we
have manifestly set the component in $\chi^3$ direction of $A$ to zero. We will
see it is sufficient for our construction.

With(\ref{pos}),(\ref{posre}),(\ref{connection}),(\ref{assign1}) the curvature
two form is easily found
\be
F\ba[t]{l} =dA(x)+A(x)\wedge A(x)\\[3mm]
	=\ba[t]{l}i(\partial_\mu A_\nu+iA_\mu A_\nu)dx^\mu\wedge dx^\nu
	-i(\partial_\mu S+i[A_\mu,S])dx^\mu\wedge \chi^1\\[3mm]
	 -i\gamma_5(\partial_\mu P+i[A_\mu,P])dx^\mu\wedge \chi^2
	 -\gamma_5[S, P]\chi^1\wedge \chi^2 \vspace{3mm}\ea\\
	=\frac{i}{2}F_{\mu\nu}dx^\mu\wedge dx^\nu
	 -iD_\mu S dx^\mu\wedge \chi^1
	 -i\gamma_5 D_\mu P dx^\mu\wedge \chi^2
	-\gamma_5[S,P]\chi^1\wedge\chi^2.
\ea
\ee
here $d=d_{Z_2\oplus Z_2}+d_M$, $d_{Z_2\oplus Z_2}$ is the external derivative
defined in the last section, $d_M$ is the usual external derivative on
Minkowski space.

Using
\be
{\cal L}=\int_{Z_2\oplus Z_2}Tr<F(h), \overline{F(h)}>
\ee
we get exactly the bosonic sector of the Lagrangian
\be
{\cal L}_B\ba[t]{ll}
=&Tr(\frac{1}{4}F_{\mu\nu}F_{\rho\sigma}<dx^\mu\wedge dx^\nu,
\overline{dx^\rho\wedge dx^\sigma}>
+D_\mu S(D_\nu S)^*<dx^\mu\wedge \chi^1,\overline{dx^\nu\wedge \chi^1}>\\[3mm]
&+D_\mu P(D_\nu P)^*<dx^\mu\wedge \chi^2,\overline{dx^\nu\wedge \chi^2}>
+[S,P][S,P]^*<\chi^1\wedge\chi^2,\overline{\chi^1\wedge\chi^2}>)\\[3mm]
=& Tr(-\frac{1}{4}F_{\mu\nu}F^{\mu\nu}
+\frac{1}{2}D_\mu SD^\mu S
+\frac{1}{2}D_\mu PD^\mu P
+\frac{1}{2}[S,P]^2)
\ea\ee
where in order to give correct normalization constants, we have chosen the
metric
\be\ba{l}
<dx^\mu\wedge dx^\nu, \overline{dx^\rho\wedge dx^\sigma}>
=<dx^\mu\wedge dx^\nu, {dx^\sigma\wedge dx^\rho}>
=g^{\mu\sigma}g^{\nu\rho}\\[3mm]
<dx^\mu\wedge\chi^1, \overline{dx^\nu\wedge\chi^1}>
=<dx^\mu\wedge\chi^1, {-\chi^1\wedge dx^\nu}>
=\frac{1}{2}g^{\mu\nu}\\[3mm]
<dx^\mu\wedge\chi^2, \overline{dx^\nu\wedge\chi^2}>
=<dx^\mu\wedge\chi^2, {-\chi^2\wedge dx^\nu}>
=\frac{1}{2}g^{\mu\nu}\\[3mm]
<\chi^1\wedge\chi^2,\overline{\chi^1\wedge\chi^2}>
=<\chi^1\wedge\chi^2,{\chi^2\wedge\chi^1}>
=\frac{1}{2}.
\ea\ee

As to the fermionic sector of the Lagrangian, we have
\be
{\cal L}_F\ba[t]{l}
=\int_{Z_2\oplus Z_2}Tr(i\bar{\psi}[\gamma^\mu D_\mu+D_1+D_2+D_3,\psi])\\[3mm]
=Tr(i\bar{\psi}\gamma^\mu D_\mu\psi
+\bar{\psi}[S+\gamma_5P, \psi])
\ea
\ee
where
\be\ba{l}
[D_1, \psi]=\partial_1\psi-i[SR_1, \psi]=-i[S, \psi]\\[3mm]
[D_2, \psi]=\partial_2\psi-i[\gamma_5PR_2, \psi]=-i[\gamma_5P, \psi]\\[3mm]
[D_3, \psi]=\partial_3\psi=0.
\ea\ee
So the $N=2$ super Yang-Mills Lagrangian is reformulated.

We next consider $N=4$ case, the Lagrangian is
\be
{\cal L}=\ba[t]{l}
Tr[-\frac{1}{4}F_{\mu\nu}F^{\mu\nu}
+i{\lambda}_i\sigma^\mu D_\mu\bar\lambda^i
+\frac{1}{2}D_\mu M^{ij} D^\mu M_{ij} \\[3mm]
+i\lambda_i[M^{ij},\lambda_j]
+i\bar{\lambda}^i[M_{ij},\bar{\lambda}^j]\\[3mm]
+\frac{1}{4}[M_{ij}, M_{kl}][M^{ij}, M^{kl}]]
\ea\ee
where  the scalar $M_{ij}$ is self-conjugate,
$M^{\dag}_{ij}=\frac{1}{2}\epsilon^{ijkl}M_{kl}=M^{ij}$, and transforms as $6$
of $SU(4)$, and $\lambda_i$ is the $2-$spinor transforming as $4$ of $SU(4)$
whereas $\bar\lambda^i$ transforming as $\bar 4$. So ${\cal L}$ is $SU(4)$
invariant.

In order to write the spinor in four component form, we introduce Majorana
spinors
\be
\lambda_i=\left(\ba{c} \lambda_{\alpha i}\\
\bar\lambda^{\dot{\alpha} i}\ea\right);\ \ \ \ \
\bar\lambda_i=\left(\ba{cc} \lambda^{\alpha}_i,&
\bar\lambda_{\dot{\alpha}}^i\ea\right)
\ee
and
\be\ba{l}
\phi_{ij}=\left(\ba{cc}M^{ij}&\\&M_{ij}\ea\right)\\[3mm]
I_{ij}=\left(\ba{cc}\delta^j_i&\\&\delta^i_j\ea\right)\\[3mm]
\ea\ee
$\alpha=1,2$ and $\dot{\alpha}=\dot{1},\dot{2}$ are spinor indices.
So the Lagrangian can be written as
\be
{\cal L}=\ba[t]{l}
Tr[-\frac{1}{8}F_{\mu\nu}F^{\mu\nu}I_2
+\frac{i}{2}\bar{\lambda}_i\gamma^\mu D_\mu\lambda_i
+\frac{1}{4}D_\mu \phi^{\dag}_{ij} D^\mu \phi_{ij} \\[3mm]
+i\bar{\lambda}_i[\phi_{ij},\lambda_j]
+\frac{1}{8}[\phi_{ij}, \phi_{kl}][\phi^{\dag}_{ij}, \phi^{\dag}_{kl}]].
\ea\ee
here the trace also takes on the $2\times 2$ matrix in which fields in two
$2-$component form are arranged into $4-$component form.

We set the connection one-form on $M^4\times(Z_2\oplus Z_2)$
\be
A(x,g)=iA_\mu(x,g)I_{ij}dx^\mu+\phi_{ij}(x,g)\chi^1+\phi_{ij}(x,g)\chi^2\ \ \ \
g\in Z_2\oplus Z_2.
\ee
Again we will let all fields take the same values on different points of
$Z_2\oplus Z_2$, so we will ignore the variable $g$, and as in the $N=2$ case,
we let the component in $\chi^3$ direction be zero.

Then we have
\be
F\ba[t]{l}=dA(x)+A(x)\wedge A(x)
=\frac{i}{2}F_{\mu\nu}I_2dx^\mu\wedge dx^\nu+D_\mu\phi_{ij}
dx^\mu\wedge\chi^1\\[3mm]
+D_\mu\phi_{ij} dx^\mu\wedge\chi^2+[\phi_{ij},\phi_{kl}]\chi^1\wedge\chi^2.
\ea\ee
here the indices $(ij)$ and $(kl)$ in the last term can not be contracted since
they belong to same representation $6$ or $\bar 6$. In the following
construction of Lagrangian we will contract the indices in $\phi$ and
$\phi^{\dag}$ because they belong to $6$ and $\bar 6$ respectively, so that
such contraction will give $SU(4)$ invariant quantity.

The bosonic sector of Lagrangian
\be
{\cal L}_B\ba[t]{ll}
=&\frac{1}{2}\int_{Z_2\oplus Z_2}Tr<F(h), \overline{F(h)}>\\[3mm]
=&\frac{1}{2}Tr(\frac{1}{4}F_{\mu\nu}F_{\rho\sigma}I_2<dx^\mu\wedge dx^\nu,
\overline{dx^\rho\wedge dx^\sigma}>
+D_\mu \phi_{ij}(D_\nu \phi_{ij})^*<dx^\mu\wedge \chi^1,\overline{dx^\nu\wedge
\chi^1}>\\[3mm]
&+D_\mu\phi_{ij}(D_\nu \phi_{ij})^*<dx^\mu\wedge \chi^2,\overline{dx^\nu\wedge
\chi^2}>
+[\phi_{ij},\phi_{kl}][\phi_{ij},\phi_{kl}]^*
<\chi^1\wedge\chi^2,\overline{\chi^1\wedge\chi^2}>)\\[3mm]
=& Tr(-\frac{1}{8}F_{\mu\nu}F^{\mu\nu}I_2
+\frac{1}{4}D_\mu \phi^{\dag}_{ij} D^\mu \phi_{ij}
+\frac{1}{8}[\phi_{ij},\phi_{kl}][\phi^{\dag}_{ij},\phi^{\dag}_{kl}])
\ea\ee
here again in order to give appropriate normalization constants, we have chosen
the metric
\be\ba{l}
<dx^\mu\wedge dx^\nu, \overline{dx^\rho\wedge dx^\sigma}>
=<dx^\mu\wedge dx^\nu, {dx^\sigma\wedge dx^\rho}>
=g^{\mu\sigma}g^{\nu\rho}\\[3mm]
<dx^\mu\wedge\chi^1, \overline{dx^\nu\wedge\chi^1}>
=<dx^\mu\wedge\chi^1, {-\chi^1\wedge dx^\nu}>
=\frac{1}{4}g^{\mu\nu}\\[3mm]
<dx^\mu\wedge\chi^2, \overline{dx^\nu\wedge\chi^2}>
=<dx^\mu\wedge\chi^2, {-\chi^2\wedge dx^\nu}>
=\frac{1}{4}g^{\mu\nu}\\[3mm]
<\chi^1\wedge\chi^2,\overline{\chi^1\wedge\chi^2}>
=<\chi^1\wedge\chi^2,{\chi^2\wedge\chi^1}>
=\frac{1}{4}.
\ea\ee

The fermionic sector of Lagrangian reads
\be
{\cal L}_F\ba[t]{l}
=\frac{1}{2}\int_{Z_2\oplus Z_2}
Tr(i\bar{\lambda_i}[\gamma^\mu
(D_\mu)I_{ij}+(D_1)_{ij}+(D_2)_{ij}+(D_3)_{ij},\lambda_j])\\[3mm]
=Tr(\frac{i}{2}\bar{\lambda_i}\gamma^\mu D_\mu\lambda_i
+i\bar{\lambda_i}[\phi_{ij}, \lambda_j])
\ea
\ee
where
\be\ba{l}
[(D_1)_{ij}, \lambda_j]= I_{ij}\partial_1\lambda_j+[\phi_{ij}, \lambda_j]
=[\phi_{ij}, \lambda_j]\\[3mm]
[(D_2)_{ij}, \lambda_j]= I_{ij}\partial_2\lambda_j+[\phi_{ij}, \lambda_j]
=[\phi_{ij}, \lambda_j]\\[3mm]
[(D_3)_{ij}, \lambda_j]= I_{ij}\partial_3\lambda_j=0.
\ea\ee
So we see the $N=4$ super Yang-Mills Lagrangian can also be reformulated on
$M^4\times (Z_2\oplus Z_2)$.

\section{Conclusions}
We show that the $N=2$ and $N=4$ super Yang-Mills Lagrangian can be
reformulated on  $M^4\times (Z_2\oplus Z_2)$ non-commutative geometry. We use
the differential calculus on discrete group developed by A.Sitarz and regard
the scalar or pseudoscalar fields as connection one-forms on the discrete
symmetry group. This reformulation is different from that of A.H.Chamseddine
and seems much simpler and clearer.

\end{document}